\documentclass[12pt,thmsa]{article}

\usepackage{dcolumn}

\usepackage{graphicx}

\begin{document}

\title{Perturbation Theory for Population Dynamics}
\author{Francisco M. Fern\'{a}ndez \\
INIFTA\ (Conicet, UNLP), Divisi\'{o}n Qu\'{i}mica Te\'{o}rica,\\
Diag. 113 y 64 (S/N), Sucursal 4, Casilla de Correo 16,\\
1900 La Plata, Argentina \\E--mail: fernande@quimica.unlp.edu.ar}

\maketitle

\begin{abstract}
We prove that a recently proposed homotopy perturbation method for the
treatment of population dynamics is just the Taylor expansion of the
population variables about initial time. Our results show that this
perturbation method fails to provide the global features of the ecosystem
dynamics.
\end{abstract}

\section{Introduction \label{sec:intro}}

Recently, Chowdhury et al \cite{CHA07} proposed the application of
homotopy--perturbation method (HPM) to simple models of population dynamics
and obtained approximate solutions in the form of perturbation series. The
authors show that their approximate analytical results agree with the
numerical solution of the problem and appear to suggest that the HPM series
are convergent.

A straightforward inspection of such series reveals that they are merely
Taylor expansions of the time variable. One does not expect such a local
approximation to provide a reasonable description of the dynamics of
nonlinear systems, except in the neighbourhood of the initial conditions.\
Singular points appear spontaneously in nonlinear systems and move around
the complex plane as the initial conditions vary \cite{BO78} which makes
unlikely that the time series are valid for all time.

The purpose of this letter is to investigate the range of utility of the
homotopy time series to provide useful insight on the dynamics of population
models. In Section \ref{sec:HPM} we show that the HPM of Chowdhury et al
\cite{CHA07} always leads to a Taylor expansions of the solution of the
nonlinear system about initial time. In Section \ref{sec:1-D} we analyze the
exactly solvable one--dimensional problem already considered by Chowdhury et
al \cite{CHA07}. In Section \ref{sec:2-D} we study their population model for
two species \cite{CHA07} and an exactly solvable two--dimensional dynamical
model. Finally, in Section \ref{sec:conclusions} we summarize our results
and draw some conclusions.

\section{Perturbation method \label{sec:HPM}}

Population models give rise to differential equations of the form
\begin{equation}
\mathbf{\dot{x}}(t)=\mathbf{f}(\mathbf{x(t)}),\;\mathbf{x}(0)=\mathbf{x}_{0}
\label{eq:nonlin}
\end{equation}
where $\mathbf{x}$ is a vector of the $n$ variables $x_{1},x_{2},\ldots
,x_{n}$ and $\mathbf{f}(\mathbf{x})$ is a vector--valued function with
components $f_{1}(\mathbf{x}),f_{2}(\mathbf{x}),\ldots ,f_{n}(\mathbf{x})$.
We assume that this vector--valued function is continuously differentiable.

The HPM proposed by Chowdhury et al \cite{CHA07} is equivalent to a
straightforward perturbation theory based on the modified equation $\mathbf{%
\dot{x}}(\lambda ,t)=\lambda \mathbf{f}(\mathbf{x}(\lambda ,t))$, where $%
\lambda $ is a dummy perturbation parameter. Notice that $\mathbf{x}(1,t)=%
\mathbf{x}(t)$ so that we set $\lambda $ equal to unity at the end of the
calculation. The HPM proposes a solution in the form of a series
\begin{equation}
\mathbf{x}(\lambda ,t)=\sum_{j=0}^{\infty }\mathbf{x}^{(j)}(t)\lambda ^{j}
\label{eq:x_lambda_series}
\end{equation}
where $\lambda $ is finally set equal to unity as said above. Notice that
since $\mathbf{\dot{x}}^{(0)}(t)=\mathbf{0}$ the resulting unperturbed or
reference model $\mathbf{x}^{(0)}(t)=\mathbf{x}_{0}$ seems to be quite poor
at first sight. Our results clearly show that it is actually the case. The
initial conditions for the perturbation corrections are $\mathbf{x}^{(j)}(0)=%
\mathbf{0}$ for all $j>0$.

If we define the new time variable $\tau =\lambda t$, then $d\mathbf{x}%
/d\tau =\mathbf{f}(\mathbf{x})$ from which we conclude that $\mathbf{x}%
(\lambda ,t)=\mathbf{x}(\lambda t)$ and that this particular implementation
of the HPM becomes the straightforward time series
\begin{equation}
\mathbf{x}(t)=\sum_{j=0}^{\infty }\mathbf{x}_{j}t^{j}  \label{eq:t_series}
\end{equation}
in agreement with the particular results derived by Chowdhury et al \cite
{CHA07}. For example, the coefficient of first order is simply $\mathbf{x}%
_{1}=\mathbf{f}(\mathbf{x}_{0})$.

\section{One--dimensional model \label{sec:1-D}}

We first consider the simple one--dimensional model \cite{CHA07}
\begin{equation}
\dot{x}=x(b+a\,x)  \label{eq:1-D}
\end{equation}
where $b\geq 0$ and $a<0$. In this case we have an unstable node at $x=0$
and a stable one at $x=-b/a$ \cite{BO78}. The exact solution is
\begin{equation}
x(t)=\left\{
\begin{array}{c}
\frac{bx_{0}e^{bt}}{b-ax_{0}(1-e^{bt})},\;b>0 \\
\frac{x_{0}}{1-ax_{0}t},\;b=0
\end{array}
\right.   \label{eq:x_1-D}
\end{equation}

As mentioned above, the HPM series agree with the Taylor expansion of the
exact solution about $t=0$:
\begin{eqnarray}
x(t) &=&x_{0}+tx_{0}(ax_{0}+b)+\frac{t^{2}x_{0}(2ax_{0}+b)(ax_{0}+b)}{2}
\nonumber \\
&&+\frac{t^{3}x_{0}(6a^{2}x_{0}^{2}+6abx_{0}+b^{2})(ax_{0}+b)}{6}  \nonumber
\\
&&+\frac{t^{4}x_{0}(2ax_{0}+b)(12a^{2}x_{0}^{2}+12abx_{0}+b^{2})(ax_{0}+b)}{%
24}+\ldots   \label{eq:t_series_1-D}
\end{eqnarray}

Equation (\ref{eq:x_1-D}) clearly sows that the solution has a pole at $%
t_{c}=b^{-1}\ln [1+b/(ax_{0})]$ and, therefore, the HPM series does not
converge for $t>|t_{c}|$. When $b=1$, $a=-3$, and $x_{0}=0.1$, then $%
|t_{c}|=3.253846656$ is much larger than the largest time value chosen by
Chowdhury et al \cite{CHA07}, which explains why those authors obtained such
good results. If, for example, $x_{0}=1$, then $t_{c}=-\ln (3/2)$ and the
HPM series is unsuitable for $t>\ln (3/2)$ as shown in Table \ref{tab:Table1}%
. Notice that these results reflect the fact that the usefulness of the HPM
depends on the initial conditions. Besides, the HPM series does not take
into consideration the stable node at $x=-b/a$ and is therefore unable to
reveal the main features of the dynamical behaviour of the system.

\section{Two--dimensional Models \label{sec:2-D}}

Chowdhury et al \cite{CHA07} also discussed the simple two--species model
\begin{eqnarray}
\dot{x} &=&x(b_{1}+a_{11}x+a_{12}y)  \nonumber \\
\dot{y} &=&y(b_{2}+a_{21}x+a_{22}y)  \label{eq:2-D_model}
\end{eqnarray}
and obtained accurate results for the model parameters $b_{1}=0.1$, $%
a_{11}=-0.0014$, $a_{12}=-0.0012$, $b_{2}=0.08$, $a_{21}=-0.0009$, $%
a_{22}=-0.001$, and initial conditions $x_{0}=4$, $y_{0}=10$. However, the
time interval considered by the authors is too small to give any indication
of the evolution of this model ecosystem.

By inspection of the first terms of the HPM expansions
\begin{eqnarray}
x(t)
&=&x_{0}+tx_{0}(a_{11}x_{0}+a_{12}y_{0}+b_{1})+t^{2}x_{0}[2a_{11}^{2}x_{0}^{2}+3a_{11}x_{0}(a_{12}y_{0}+b_{1})+a_{12}^{2}y_{0}^{2}
\nonumber \\
&&+a_{12}y_{0}(a_{21}x_{0}+a_{22}y_{0}+2b_{1}+b_{2})+b_{1}^{2}]/2+\ldots
\nonumber \\
y(t)
&=&y_{0}+ty_{0}(a_{21}x_{0}+a_{22}y_{0}+b_{2})+t^{2}y_{0}[a_{11}a_{21}x_{0}^{2}+a_{12}a_{21}x_{0}y_{0}+a_{21}^{2}x_{0}^{2}
\nonumber \\
&&+a_{21}x_{0}(3a_{22}y_{0+}b1+2b_{2})+2a_{22}^{2}y_{0}^{2}+3a_{22}b_{2}y_{0}+b_{2}^{2}]/2+\ldots
\label{eq:2-D_time_series}
\end{eqnarray}
we appreciate that if the model parameters $b_{i}$ and $a_{ij}$ are
sufficiently small (as those chosen by Chowdhury et al \cite{CHA07}), then
the time series may give accurate results for an apparently large time
interval.\ However, these series do not take into consideration the critical
points of the model equations and therefore they cannot reveal the actual
dynamics of the system \cite{BO78,S94}.

The nonlinear dynamical system (\ref{eq:2-D_model}) exhibits four critical
points in phase space \cite{BO78,S94}: an unstable node at $(x,y)=(0,0)$,
two saddle points at $(x,y)=(0,80)$ and $(x,y)=(71.43,0)$, and a stable node
at $(x,y)=(12.5,68.75)$.

\begin{figure}[H]
\begin{center}
\includegraphics[width=9cm]{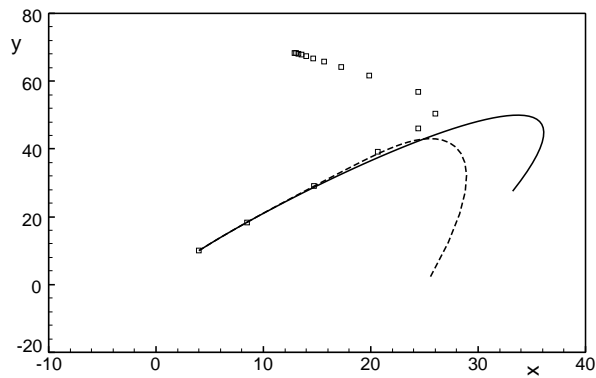}
\end{center}
\caption{Numerical (squares), fourth--order time series (solid) and
tenth--order time series (dashed) curves in the phase plane for model (\ref{eq:2-D_model}). }
\label{fig:2-D}
\end{figure}

Fig. \ref{fig:2-D} shows that the population moves in phase space from the
initial condition to the stable node and that the time series is unable to
take into account this important dynamical behaviour. Increasing the
perturbation order from four (the one used by Chowdhury et al \cite{CHA07})
to ten just improves the accuracy for small time but worsens it at larger
time which suggests that the convergence radii of the time series are rather
too small. In other words, the time series do not allow us to study the
important population portrait in phase space.

In order to appreciate a more dramatic failure of the time series consider
the following system of nonlinear equations \cite{S94}
\begin{eqnarray}
\dot{x} &=&-y+ax(x^{2}+y^{2})  \nonumber \\
\dot{y} &=&x+ay(x^{2}+y^{2})  \label{eq:2-D_solvable}
\end{eqnarray}
It is unsuitable for population dynamics because it allows negative values
of $x(t)$ and $y(t)$ but has the great advantage of being exactly solvable.
Its solutions are
\begin{eqnarray}
x(t) &=&\frac{r_{0}\cos (\theta _{0}+t)}{\sqrt{1-2ar_{0}^{2}t}},\;y(t)=\frac{%
r_{0}\sin (\theta _{0}+t)}{\sqrt{1-2ar_{0}^{2}t}}  \nonumber \\
r_{0} &=&\sqrt{x_{0}^{2}+y_{0}^{2}},\;\theta _{0}=\arctan \left( \frac{y_{0}%
}{x_{0}}\right)   \label{eq:2-D_sols}
\end{eqnarray}
We have an unstable spiral when $a>0$ and a stable one when $a<0$ \cite{S94}%
. We clearly see the pole at $t_{c}=1/(2ar_{0}^{2})$ and realize that the
time series will be completely useless for $t>|t_{c}|$.

\begin{figure}[H]
\begin{center}
\includegraphics[width=9cm]{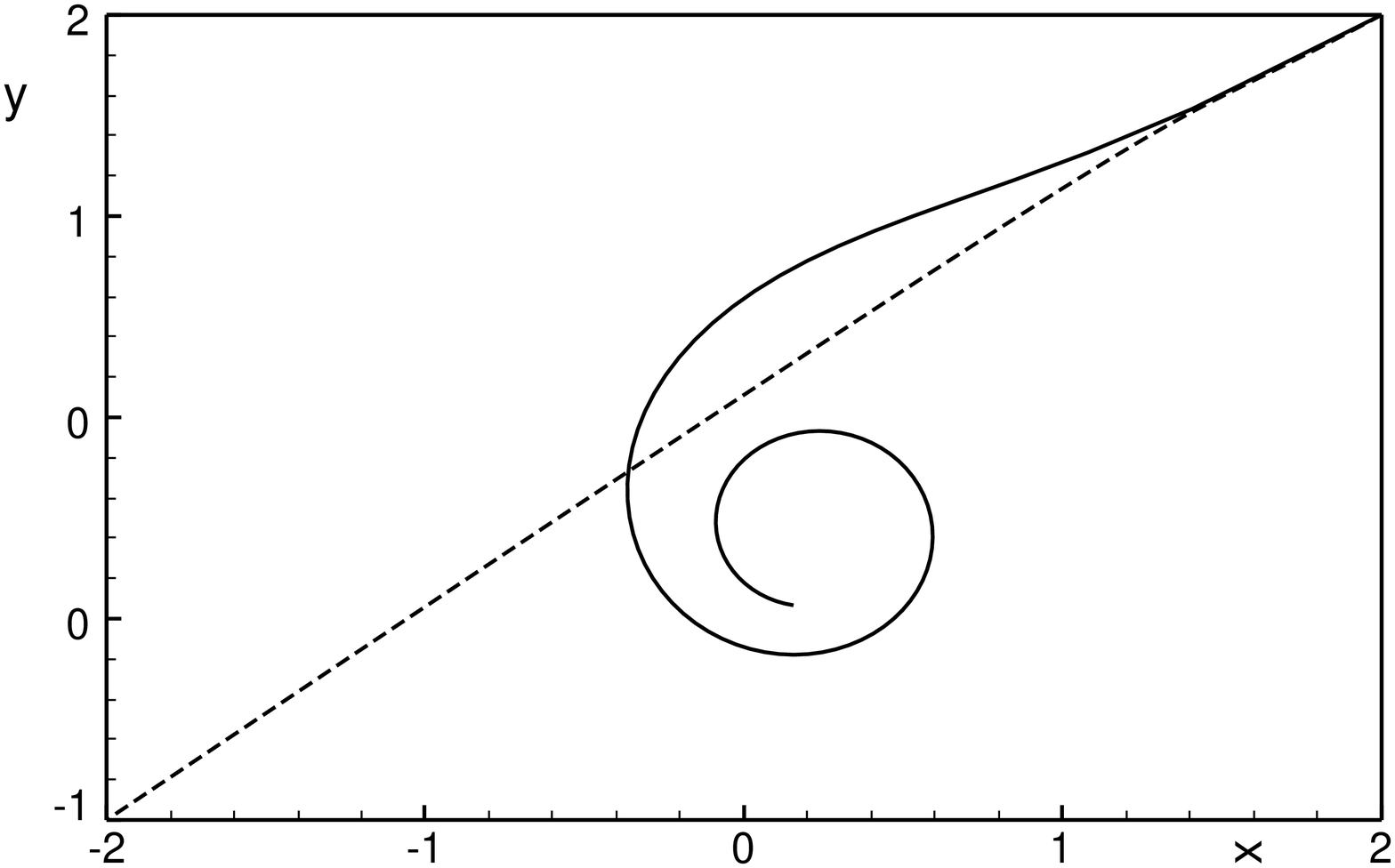}
\end{center}
\caption{Numerical (solid) and fifth--order time series (dashed) curves in
the phase plane for model (\ref{eq:2-D_solvable}). }
\label{fig:2-D_exact}
\end{figure}

Fig. \ref{fig:2-D_exact} shows the numerical and approximate solutions for
this model when $a=-0.5$ and $x_{0}=y_{0}=2$. We clearly notice that the
time series fail completely to provide a qualitative description of the
spiral point and thereby of the global details of the system dynamics.

It is worth mentioning that multiple--scale perturbation theory \cite
{BO78,S94} gives the exact answer for this model and therefore appears to be
a much more reliable perturbation approach for nonlinear dynamics.

\section{Conclusions \label{sec:conclusions}}

We have shown that:

\begin{itemize}
\item  the homotopy perturbation method proposed by Chowdhury et al \cite
{CHA07} is just the Taylor expansion of the solutions of the nonlinear
systems about $t=0$.

\item  the perturbation series, and consequently the HPM, is limited to a
neighbourhood of the initial time determined by the singular point closest
to the origin of the complex $t$--plane. The locations of the singular
points of the nonlinear models shift as the initial conditions vary.

\item  the HPM does not give an acceptable qualitative description of the
most important features of the evolution of the dynamical system in phase
space. In this sense, the HPM is by far less useful than the standard
linearization which is also a local approach \cite{BO78,S94}.
\end{itemize}

In principle, other implementations of the homotopy perturbation method may
be more suitable for nonlinear dynamics. For example, we may choose the
linear approximation about the critical or fixed points \cite{BO78,S94} as
unperturbed or reference model for the subsequent application of
perturbation theory.

Since homotopy perturbation methods have become quite popular and are
currently being applied to a wide variety of fields \cite{CHA07,H06} (and
references therein), present results become important because they suggest
that a more careful scrutiny of the approach's performance is required.

\begin{table}[H]
\caption{Logarithmic error $\log |(Exact-Approximate)/Exact|$ for the
fourth--order time series for model (\ref{eq:1-D})}
\label{tab:Table1}
\begin{center}
\par
\begin{tabular}{cD{.}{.}{10}}  \hline
$t$ & \multicolumn{1}{c}{Logarithmic Error} \\ \hline

 0.1 & -3.14   \\
 0.2 & -1.66   \\
 0.3 & -0.798  \\
 0.4 & -0.193  \\
 0.5 & 0.273   \\
 0.6 & 0.651   \\
 0.7 & 0.968   \\
 0.8 &  1.24   \\
 0.9 &  1.48   \\
 1.0 &  1.69

\end{tabular}
\end{center}
\end{table}

\begin{thebibliography}{9}
\bibitem{CHA07}  M. S. H. Chowdhury, I. Hashim, and O. Abdulaziz, Phys.
Lett. A 368 (2007) 251.

\bibitem{BO78}  C. M. Bender and S. A. Orszag, Advanced mathematical methods
for scientists and engineers, (McGraw-Hill, New York, 1978).

\bibitem{S94}  S. H. Strogatz, Nonlinear Dynamics and Chaos, with
Applications to Physics, Biology, Chemistry, and Engineering, (Perseus
Books, Reading, Massachusetts, 1994).

\bibitem{H06}  J. H. He, Int. J. Mod. Phys. B 20 (2006) 1141.
\end{thebibliography}
\end{document}